\documentclass[10pt,a4paper]{article}
\usepackage[latin1]{inputenc}
\usepackage[affil-it]{authblk}
\usepackage{amsmath}
\usepackage{amsfonts}
\usepackage{amssymb}
\usepackage{cases}
\usepackage{dsfont}
\usepackage{color}
\usepackage{graphicx}
\usepackage{epsfig}
\usepackage{bm}
\usepackage{bbm}
\usepackage{ulem}
\usepackage{enumerate}
\usepackage{url}
\usepackage{hyperref}

\newtheorem{theorem}{Theorem}[section]

\newcommand\seq[3]{{#1}_{#2}^{#3}}
\newcommand{\bd}{\begin{document}}
\newcommand{\ed}{\end{document}}
\newcommand{\beq}{\begin{equation}}
\newcommand{\eeq}{\end{equation}}
\newcommand{\bef}{\begin{figure}}
\newcommand{\enf}{\end{figure}}
\newcommand{\bea}{\begin{eqnarray}}
\newcommand{\eea}{\end{eqnarray}}
\newcommand{\baR}{\begin{array}}
\newcommand{\eaR}{\end{array}}
\newcommand{\bc}{\begin{center}}
\newcommand{\ec}{\end{center}}
\newcommand{\ben}{\begin{enumerate}}
\newcommand{\een}{\end{enumerate}}
\newcommand{\bit}{\begin{itemize}}
\newcommand{\eit}{\end{itemize}}
\newcommand{\su}{\section}
\newcommand{\ssu}{\subsection}
\newcommand{\sssu}{\subsubsection}
\newcommand\ssssu[1]{\textbf{#1}}
\newcommand{\nid}{\noindent}
\newcommand{\nnb}{\nonumber}
\newcommand\bloc[2]{{\omega}_{#1}^{#2}}
\newcommand\blocp[2]{{\bra{\sigma \, \omega}}_{#1}^{#2}}
\newcommand{\un}{\mathbbm{1}}
\newcommand{\setZ}{\mathbbm{Z}}
\newcommand{\setN}{\mathbbm{N}}
\newcommand{\setR}{\mathbbm{R}}
\newcommand{\pare}[1]{\left(\, #1 \, \right)}
\newcommand{\scal}[1]{\langle\, #1 \, \rangle}
\newcommand{\bra}[1]{\left[\, #1 \, \right]}
\newcommand{\brac}[2]{\left[\, #1 \, | \, #2 \right]}
\newcommand{\Set}[1]{\left\{\, #1 \, \right\}}
\newcommand{\Setc}[2]{\left\{\, #1 \, ; \, #2 \right\}}
\newcommand{\Prob}[2]{\mathds{P}_{#1}\left[\, #2 \, \right]}
\newcommand{\PT}[2]{\pi^{(T)}_{#2}\left[\, #1 \, \right]}
\newcommand{\pT}{\pi^{(T)}}
\newcommand{\Probex}[1]{P^{(\ast)}\left[\, #1 \, \right]}
\newcommand{\Probc}[3]{\mathds{P}_{#1}[#2 \, \left| \, #3 \right.]}
\newcommand{\Pnc}[2]{mathds{P}_n\bra{#1 \, \left| \, #2 \right.}}
\newcommand{\Probcex}[2]{P^{(\ast)}\bra{#1 \, \left| \, #2 \right.}}
\newcommand{\Probcp}[2]{P^{(ap)}\bra{#1 \, \left| \, #2 \right.}}
\newcommand{\abs}[1]{\left|\, #1 \, \right|}
\newcommand{\deq}{\stackrel {\rm def}{=}}
\newcommand{\sqsubsetneq}{\stackrel {\tiny\sqsubset}{\tiny \neq}}
\newcommand\moy[1]{\mu\bra{#1}}
\newcommand\cA{{\cal A}}
\newcommand\cB{{\cal B}}
\newcommand\cC{{\cal C}}
\newcommand\cD{{\cal D}}
\renewcommand\H{{\cH_\bbeta}}
\renewcommand{\sb}{s_\bbeta}
\newcommand{\sB}{\sigma_B}
\newcommand\cF{{\cal F}}
\newcommand\cG{{\cal G}}
\newcommand\cI{{\cal I}}
\newcommand\cH{{\cal H}}
\newcommand\Q{\mathcal{Q}}
\newcommand\cK{{\cal K}}
\newcommand\cL{{\cal L}}
\newcommand\cM{{\cal M}}
\newcommand\cT{{\cal T}}
\newcommand\cS{{\cal S}}
\newcommand\cP{{\cal P}}
\newcommand\cO{{\cal O}}
\newcommand\cU{{\cal U}}
\newcommand\cV{{\cal V}}
\newcommand\cW{{\cal W}}
\newcommand\cX{{\cal X}}
\newcommand\f{{\bm{f}}}
\newcommand\bF{{\bm{F}}}
\newcommand\bo{{\bm{o}}}
\newcommand\blambda{{\bm{\lambda}}}
\newcommand\bbeta{{\bm{\beta}}}
\newcommand\gk[1]{g_k\pare{#1}}
\newcommand\gLk{g_{L,k}}
\newcommand\akj[1]{\alpha_{kj}(#1)}
\newcommand\sif[1]{\seq{\omega}{-\infty}{#1}}
\newcommand\Xk[1]{X_k({#1})}
\newcommand\Ik[1]{I_k({#1})}
\newcommand\Skj[1]{S_{kj}({#1})}
\newcommand\XkD{\Xk{\bloc{0}{D-1}}}
\newcommand\Xkr{X_k^r}
\newcommand\tLk{\tau_{L,k}}
\newcommand\moyc[2]{\mu\bra{#1 \, | \, #2}}
\newcommand\Vkdet[1]{V_k^{(det)}({#1})}
\newcommand\Vksyn[1]{V_k^{(syn)}({#1})}
\newcommand\Vkext[1]{V_k^{(ext)}({#1})}
\newcommand\Vknoise[1]{V_k^{(noise)}({#1})}
\newcommand\sk[1]{\sigma_k({#1})}
\newcommand\skr{\sigma_k^r\pare{\bloc{0}{D-1}}}
\newcommand{\ent}[1]{\left[\, #1 \, \right]}
\newcommand\vm[1]{var_m\left[#1 \right]}
\newcommand\eg[1]{\stackrel {\rm #1}{=}}
\newcommand\tk[1]{\tau_k\pare{#1}}
\newcommand\tko{\tau_k(t,\omega)}
\newcommand\cBm[1]{\cB^{-1}\pare{#1}}
\newcommand{\bth}{\begin{theorem}}
\newcommand{\enth}{\end{theorem}}
\newcommand\omD[1]{\seq{\bra{\omega^{(#1)}}}{0}{D-1}}
\newcommand\omz[1]{\omega^{(#1)}(D)}
\newcommand\om[1]{\omega^{(#1)}}
\newcommand\Om[1]{\Omega^{(#1)}}
\newcommand\skjq{s_{kj;q}}
\newcommand\ikq{i_{k;q}}
\newcommand\skjqs{p_{kj;q_s}}
\newcommand\ikqs{i_{k;q_s}}
\newcommand\ikqsu{i_{k;q_{s_u}}}
\newcommand\ikqu{i_{k;q_u}}
\newcommand\xkq{x_{k;q}}
\newcommand\skjqv{s_{kj;q_{v}}}
\newcommand\tjr{t_j^{(r)}(\omega)}
\newcommand\dtt{\Phi(t,\omega)}
\newcommand\dts{\Phi(s,\omega)}
\newcommand\dto{e^{\Phi(t-t_0)}}
\newcommand\dtes{e^{\Phi(t-s)}}

\newcommand\rpf{R}
\newcommand\rpfc[1]{R\pare{#1}}
\newcommand\lpf{L}
\newcommand\lpfc[1]{L\pare{#1}}
\newcommand{\zb}{\zeta_{\bbeta}}
\newcommand{\etab}{\eta_{\bbeta}}
\newcommand{\mex}{\mu^{(\ast)}}
\newcommand{\phex}{\phi^{(\ast)}}
\newcommand{\Phex}{\Phi^{(\ast)}}
\newcommand{\Hex}{\cH_{\bbeta'}^{(\ast)}}
\newcommand\p[1]{\cP\bra{#1}}
\newcommand\w[2]{w_{#1}^{(#2)}}
\newcommand\Gb{\cG_\bbeta}
\newcommand\blp[3]{\omega^{#3,(#1)}_{#2}}
\newcommand\blpb[3]{\overline{\omega^{#3,(#1)}_{#2}}}
\newcommand\moyex[1]{\mu^{(\ast)}\bra{#1}}
\newcommand\E[1]{{\cal E}_{#1}}
\newcommand\Gap{\overline{G}}
\newcommand\gap[1]{\overline{g_{#1}}}
\newcommand{\K}[2]{\cK^{(#1)}(#2)}
\newcommand{\tO}[3]{\tau^{(#1)}(#2,#3)}
\newcommand{\Vr}{V_{reset}}
\newcommand{\bcoul}[2]{{\color{#1}{\textbf{#2}}}}
\newcommand{\bruno}[1]{\bcoul{red}{#1}}
\newcommand{\rod}[1]{\bcoul{blue}{#1}}
\newcommand{\tf}[2]{t_{#1}^{(#2)}}
\newcommand{\Tf}[1]{\mathcal{T}^{(#1)}(\omega)}
\newcommand{\Exp}[1]{\mathds{E}\left[\, #1 \, \right]}
\newcommand{\Expc}[2]{\mathds{E}\left[\, #1 \, | #2 \, \right]}
\newcommand{\Cov}[1]{Cov\left[\, #1 \, \right]}
\newcommand{\bD}[2]{\underline{D}_{#1}(#2,\omega)}
\newcommand{\hk}{\hat{k}}

\begin{document}

\title{Spike train statistics and Gibbs distributions}

\author{B. Cessac\thanks{\texttt{email:bruno.cessac@inria.fr} 
}}  \author{R. Cofr\'e\thanks{\texttt{Corresponding author, email: rodrigo.cofre\textunderscore torres@inria.fr, Postal address: 2004 Route des Lucioles, 06902 Sophia-Antipolis, France.
Phone: +33-4-9238-2420, Fax: +33-4-9238-7845}}}
\affil{NeuroMathComp team (INRIA, UNSA LJAD) 2004 Route des Lucioles, 06902 Sophia-Antipolis, France.}
\date{}
\maketitle
\begin{abstract}
This paper is based on a lecture given in the LACONEU summer school, Valparaiso, January 2012. We introduce Gibbs distribution in a general setting, including non stationary dynamics, and present then three examples of such Gibbs distributions, in the context of neural networks spike train statistics: (i) Maximum entropy model with spatio-temporal constraints; (ii) Generalized Linear Models; (iii) Conductance based Integrate and Fire model with chemical synapses and gap junctions.
\end{abstract}
\section{Introduction}

\quad Neurons communicate among them by generating action potentials or ``spikes" which are pulses of electrical activity. 
When submitted to external stimuli, sensory neurons produce sequences of spikes or ``spike trains" constituting a collective response and a dynamical way to encode information about those stimuli. However, neural responses are typically not exactly reproducible, even for repeated presentation of a fixed stimulus. Therefore, characterizing the relationship between sensory stimuli and neural spike responses can be framed as a problem of determining the most adequate probability distribution  relating a stimulus to its neural response. There exist several attempts to infer this probability from data and / or general principles, based on Poisson or more general point processes \cite{pillow-ahmadian-etal:11,daley-vare:03,snyder-miller:91}, Bayesian approaches \cite{koyoma-etal:10,gerwinn-etal:09}, maximum entropy \cite{schneidman-berry-etal:06,vasquez-marre-etal:12} (for a review see \cite{rieke-etal:96}). In this paper we present several situations where the notion of Gibbs distributions is appropriate to address this problem. 

 The concept of Gibbs distribution comes from statistical physics. We use it here in a more general sense than the one usually taught in standard physics courses, although it is part of mathematical statistical physics \cite{georgii:88}. 
We argue here that Gibbs distributions might be canonical models for spike train statistics analysis. This  statement is based on three prominent examples.

\begin{enumerate}
\item The so-called \textit{Maximum Entropy Principle} allows one to propose spike train statistics models considering restrictions based on empirical observations. Although this approach has been initially devoted to show the role of weak instantaneous pairwise correlations in the retina  \cite{schneidman-berry-etal:06}, it has been recently applied to investigate the role of more complex events such as instantaneous triplets \cite{ganmor-segev-etal:11a} or spatio-temporal events \cite{vasquez-marre-etal:12}.  Probability distributions arising from the Maximum Entropy Principle are Gibbs distributions. 

\item Other approaches such as the Linear-Non Linear (LN) or Generalized Linear Models (GLM) propose an ad hoc form for the conditional probability that a neuron fires given the past network activity and given the stimulus. Those models have been proven quite efficient for retina spike trains analysis \cite{pillow-ahmadian-etal:11b}. They are not limited by the constraint of stationarity, but they are based on a questionable assumption of \textit{conditional independence between neurons}. As we show, the probability distributions coming out from those models are also Gibbs distributions.

\item Recent investigations on neural networks models (conductance based integrate-and-fire (IF) with chemical and electric synapses) show that statistics of  spike trains generated by these models are Gibbs distributions reducing to 1 when dynamics is stationary, and reducing to 2 in specific cases \cite{cessac:11a,cessac:11b,cofre-cessac:12}. In the general case, the spike trains produced by these models have Gibbs distributions \textit{which neither match 1 nor 2}. 

\end{enumerate}

The paper is organized as follows. After some definitions regarding spike train statistics and a presentation of Gibbs distributions we develop these three examples, with a short discussion of their advantages and drawbacks in spike trains analysis. Then, we discuss some relations between these models, mainly based on the Hammersley-Clifford theorem \cite{hammersley-clifford:71,besag:74,moussouris:74,clifford:90}.  This paper is a summary of several papers written by the authors and other collaborators \cite{cessac:11a,cessac:11b,nasser-marre-etal:12,cessac-palacios:12,cofre-cessac:12}. As such it does not contain original material (except the presentation).
\section{Definitions}

\subsection{Spike trains}

We consider a network of $N$ neurons.
We  assume that there is a minimal time scale $\delta > 0$ corresponding to 
the minimal resolution of the spike time, constrained  by biophysics and by measurements methods (typically $\delta \sim 1 \, ms$)
\cite{cessac-vieville:08,cessac-paugam-moisy-etal:10}. Without loss of generality (change of time units)
we set $\delta=1$,
 so that spikes are recorded at integer times.
One then associates to each
neuron $k$ and each integer time $n$ a variable $\omega_k(n)=1$ if neuron $k$ fires at time $n$ and  $\omega_k(n)=0$
otherwise.  A \textit{spiking pattern} is a vector  $\omega(n) \deq \bra{\omega_k(n)}_{k=1}^{N}$
which tells us which neurons are firing at time $n$. 
We note $\cA=\Set{0,1}^N$ the set of spiking patterns.  
A \textit{spike block} is a finite ordered list of spiking patterns, written:
$$\bloc{n_1}{n_2} = \Set{\omega(n)}_{\{n_1 \leq n \leq n_2\}},$$
where spike times have been prescribed between the times $n_1$ to $n_2$ (i.e. , $n_2-n_1+1$ time steps). 
The \textit{range} of a block is $n_2-n_1+1$, the number of time steps from $n_1$ to $n_2$.
The set of such blocks is
 $\cA^{n_2-n_1+1}$. Thus, there are $2^{Nn}$ possible blocks with $N$ neurons and range $n$. 
We call a \textit{raster plot} 
 a bi-infinite sequence $\omega \deq \left\{\omega(n)\right\}_{n=-\infty}^{+\infty}$,
of spiking patterns. Obviously experimental rasters are finite, but the consideration of infinite sequences is more convenient mathematically.  The set of raster plots is denoted $\Omega = \cA^\setZ$.

\subsection{Transition probabilities}\label{SPTrans}
The probability that a neuron emits a spike at some time $n$
depends on the history of the neural network.
However,  it is impossible to know explicitly its form in the general case
 since it depends on the past evolution of all  variables 
determining the neural network state. A possible simplification is to consider that this probability depends \emph{only} on the spikes emitted in the past by the network. In this way,
we are seeking a family of transition probabilities of the form
$\Probc{n}{\omega(n)}{\bloc{n-D}{n-1}}$, the probability
that the firing pattern $\omega(n)$ occurs at time $n$, given
a past spiking sequence $\bloc{n-D}{n-1}$. Here, $D$ is the \textit{memory depth} 
 of the probability, i.e., how
far in the past does the transition probability depend on the past spike sequence. 
We use  the convention that $\Probc{n}{\omega(n)}{\bloc{n-D}{n-1}}=\Prob{n}{\omega(n)}$
if $D=0$ (memory-less case). 

The index $n$ of $\Probc{n}{.}{.}$ indicates that transition probabilities depend explicitly on the time $n$.
We say that those transition probabilities are \textit{time-translation invariant} or \textit{stationary} if for all $n$,
$\Probc{n}{\omega(n)}{\bloc{n-D}{n-1}}=\Probc{D}{\omega(D)}{\bloc{0}{D-1}}$ whenever $\bloc{n-D}{n-1} = \bloc{0}{D-1}$ (i.e the probability does not depend explicitely on time).
In this case  we drop the index $n$.

Transition probabilities  depend on the neural network characteristics such as neurons conductances, synaptic responses or external currents. They give information about the dynamics that takes place in the observed neural network. Especially, they have a \textit{causal} structure where the probability of an event depends on the past. This reflects underlying biophysical mechanisms in the neural network, which are also causal.

\subsection{Gibbs distribution}\label{SGibbsdef}

We define here Gibbs distributions (or Gibbs measures) in a more general setting that the one usually taught in statistical physics courses, where Gibbs distributions are considered in the realm of stationary process and maximum entropy principle. Here, we do not assume stationarity and the definition encompasses the maximum entropy distributions. The Gibbs distributions considered here are called \textit{chains with complete connections} in the realm of stochastic processes \cite{fernandez-maillard:05,maillard:07} and \textit{g-measures} in ergodic theory \cite{keane:72}. They are also studied in mathematical statistical physics \cite{georgii:88}.

\sssu{Continuity with respect to a raster}\label{SCont}

For $n \in \setZ$, we note $\seq{\cA}{-\infty}{n-1}$ the set of sequences $\sif{n-1}$. Assume that we are given a set of transitions probabilities, like in the previous section, possibly depending on an infinite past\footnote{In this case, one has to assume that (i) for every $\omega(n) \in \cA$ ,
$\Probc{n}{\omega(n)}{.}$ is measurable with respect to $\cF_{\leq n-1}$, the sigma-algebra on $\seq{\cA}{-\infty}{n-1}$; (ii) for every $\sif{n-1} \in \seq{\cA}{-\infty}{n-1}$, $ \sum_{\omega(n) \in \cA} \Probc{n}{\omega(n)}{\sif{n-1}}=1$.}, i.e. of the form $\Probc{n}{\omega(n)}{\sif{n-1}}$. We give in section \ref{SGIF} an example of neural network model where such transition probabilities with an infinite memory do occur.

Even if transition probabilities involve an infinite memory $\sif{n-1}$, it is reasonable to consider situations where the effects of past spikes decreases exponentially with their distance in the past. This corresponds to the mathematical notion of \textit{continuity with respect to a raster}.
We note, for $n \in \setZ$, $m \geq 0$, and $r$ integer:
$$\omega \eg{m,n} \omega' \quad \mbox{if} \quad\omega(r)=\omega'(r), \, \forall r \in \Set{n-m, \dots, n}.
$$
Consider a function $f$ depending both on discrete time $n$ and on the raster part of $\omega$ \textit{anterior to $n$}. We write $f(n,\omega)$ instead of $f(n,\bloc{-\infty}{n-1})$.
The function $f$ is \textit{continuous with respect to the raster $\omega$} if its $m$-\textit{variation}:
\beq\label{Variation}
\vm{f(n,.)}:=\sup\Set{ \, | \,f(n,\omega)-f(n,\omega')\, | \,: \omega \eg{m,n} \omega' }
\eeq
tends to $0$ as $m \to +\infty$. This precisely means that the effect, on the value of $f$ at time $n$, as this change is more distant in the past.

\sssu{Gibbs distribution}

\nid\textbf{Definition 2.1}
A Gibbs distribution is a probability measure $\mu: \Omega \rightarrow [0,1]$ such that:
\begin{enumerate}[(i)]

\item for all $n \in \setZ$ and all $\cF_{\leq n}$-measurable functions $f$:
$$\int f\left(\seq{\omega}{-\infty}{n}\right) \mu(d\omega) = \int \sum_{\omega(n) \in \cA}
f\left(\sif{n-1} \omega(n) \right) \Probc{n}{\omega(n)}{\sif{n-1}} \mu(d\omega).$$
\item $\forall n  \in \setZ$, $\forall \sif{n-1} \in \seq{\cA}{-\infty}{n-1}$, $\Probc{n}{\omega(n)}{\sif{n-1}}>0$.
\item  For each $n \in \setZ$, $\Probc{n}{\omega(n)}{\sif{n-1}}$ is continuous with respect to $\omega$.
\end{enumerate}

The condition (i) is a natural extension of the condition
defining the invariant probability of an homogeneous Markov chain (see eq. (\ref{Gibbs_stat}) next section). In its most general sense (i) does not require stationarity and affords the consideration of an infinite memory. It defines so-called \textit{compatibility conditions}. They state that the average of a function $f(n,\omega)$ with respect to $\mu$, at time $n$ (left hand side), is equal to the average computed from transition probabilities (right hand side). This equality must hold for any time $n$.

There exist several theorems guaranteeing the existence and uniqueness of a Gibbs distribution \cite{georgii:88,fernandez-maillard:05}: this holds if the variation of transition probability decays sufficiently fast with time (typically exponentially) as $n-m \to -\infty$.

\subsection{Markov chains}\label{SMarkov}

Straightforward examples of Gibbs distributions defined that way are provided by Markov chains with positive transition probabilities. Recall that a Markov chain of length $D$ is defined by a set of transition probabilities $\Probc{n}{\omega(n)}{\bloc{n-D}{n-1}}$ where the memory depth $D>0$ is finite. These transition probabilities are obviously continuous with respect to $\omega$. If we assume moreover that they are strictly positive  $\forall n  \in \setZ$, $\forall \sif{n-1} \in \seq{\cA}{-\infty}{n-1}$ then they match (ii) in the definition above. Finally, in this case, (i) is equivalent to the following property. For any time $n_1, \, n_2, n_2-n_1 \geq D$:
\beq\label{Gibbs_stat}
\moy{\bloc{n_1}{n_2}} 
\, = \, \prod_{l=n_1+D}^{n_2} \Probc{l}{\omega(l)}{\bloc{l-D}{l-1}} \, \moy{\bloc{n_1}{n_1+D-1}}.
\eeq
For any times $n_1, n_2$ as above, the Gibbs-probability $\moy{\bloc{n_1}{n_2}}$ is given by\footnote{One also says that $\mu$ is compatible with the set of transition probabilities.} the product of the Gibbs probability of the ``initial block" $\moy{\bloc{n_1}{n_1+D-1}}$ and the products of transition probabilities from the initial time $n_1+D$ to the last time $n_2$.

Here we have considered transition probabilities depending explicitly on time $n$. When they are time-translation invariant (homogeneous Markov chain) the definition (2.1)
is the definition of the unique invariant distribution of the Markov chain (it is unique because we have assumed positive transition probabilities).\\

Let us now state (\ref{Gibbs_stat}) in a different form. Define:
\beq\label{phin}
\phi_n\pare{n,\omega} := \log \Probc{n}{\omega(n)}{\bloc{n-D}{n-1}},
\eeq
called a \textit{normalized Gibbs potential}. Then, (\ref{Gibbs_stat}) can be stated using:
\beq\label{pCondphi}
\mu\bra{\seq{\omega}{n_1}{n_2} \, | \, \bloc{n_1}{n_1+D-1}}= \exp{\sum_{l=n_1+D}^{n_2} \phi_l\pare{l,\omega}}.
\eeq

This form
reminds the Gibbs distribution on spin lattices in statistical physics where one looks for
lattice translation-invariant probability distributions given specific boundary conditions.
Given a  potential of range $D$ the probability of a spin block depends on the states of spins
in a neighborhood of size $D$ of that block. Thus,  the conditional probability of this block
given a fixed neighborhood is the exponential of the energy characterizing physical interactions within the block as well as with the boundaries. 
Here, spins are replaced by spiking patterns;
space is replaced with time which is mono-dimensional and oriented: there is no dependence in the future.
Boundary conditions are replaced
by the dependence in the past.\\

The definition (\ref{phin}) of the normalized Gibbs potential 
extends to the case $D \to +\infty$.

\section{Gibbs distributions and models of spike train statistics}

In this section we review several examples of models/concepts used to analyze spike train statistics. All of them enter in the realm of Gibbs distributions defined above.

\subsection{Maximum entropy models}\label{SMaxEnt}

The definition (2.1) affords time-dependent transition probabilities. On the opposite, in this section we assume that they do not depend explicitly on $n$, or, equivalently, that they are time-translation invariant. This corresponds to the physical concept of \textit{stationarity}.  

Assume that spike trains statistics is distributed according to an hidden probability $\mu$. How to approach $\mu$ from data ? Maximum entropy  provides a method that allows to approach $\mu$. It selects among all the probability distributions consistent with empirical data constraints, the most random i.e. the one with the highest entropy. But, why should we choose the maximum entropy distribution? The answer is that since entropy is a measure of information, then one should choose the probability that includes the least amount of information we have about the system and no more. The result  probability is a Gibbs distribution.

\subsubsection{Entropy}
We define  the \textit{entropy rate} (or Kolmogorov-Sinai entropy) of a probability $\mu \in \cM_{inv}$ the set of time-translation invariant probability measures as:
\beq\label{Stat_Ent}
h\bra{\mu} \, = \, - \, \limsup_{n \to \infty} \frac{1}{n+1} \, \sum_{\bloc{0}{n}} \, \moy{\bloc{0}{n}} \, \log \moy{\bloc{0}{n}},
\eeq
where the sum holds over all possible blocks $\bloc{0}{n}$. Note, that in the case of a Markov chain $h\bra{\mu}$ also reads \cite{cornfeld-fomin-sinai:82}:
\beq\label{Ent_Markov}
h\bra{\mu} \, = \, - 
 \sum_{\bloc{0}{D}}
 \, \moy{\bloc{0}{D}} \,
  \Probc{}{\omega(D)}{\bloc{0}{D-1}} \,
 \log 
 \Probc{}{\omega(D)}{\bloc{0}{D-1}}, 
 \eeq
Finally,  when $D=0$, $h\bra{\mu}$ reduces to the  usual definition:

\beq\label{entropie_spatial}
h(\mu)=-\sum_{\omega(0)} \moy{\omega(0)} \log \moy{\omega(0)}.
\eeq

We used here the notation $h(\mu)$ instead of $S$ or $s$, used in statistical physics. This is the conventional notation in ergodic theory for the (Kolmogorov-Sinai) entropy where the dependence on the measure $\mu$ is made explicit. 
\sssu{Observables}
We call \textit{observable} a function:

\begin{eqnarray}\label{DefObs}
\cO: \Omega &\rightarrow & \{0,1\}, \notag \\
\omega &\mapsto & \prod_{u=1}^r \omega_{k_u}(n_u) 
\end{eqnarray}
%
i.e. a product of binary spike events where $k_u$ is a neuron 
index and $n_u$ a time index, with $u=1, \dots, r$, for some integer $r>0$.
 Typical choices of observables are
$\omega_{k_1}(n_1)$ which is $1$ if neuron $k_1$ fires at time $n_1$ and is $0$ otherwise;
$\omega_{k_1}(n_1) \, \omega_{k_2}(n_2)$  which is $1$  if neuron $k_1$ fires at time $n_1$ and neuron $k_2$ fires at time $n_2$ and is $0$ otherwise. Another
example is $\omega_{k_1}(n_1) \,(1-\omega_{k_2}(n_2))$ which is $1$ is neuron $k_1$ fires at time $n_1$ \textit{and neuron $k_2$ is silent at time $n_2$}. This 
example emphasizes that observables are able to consider events where some neurons are silent.

We say that an observable $\cO$ has \textit{range $R$} if it depends on $R$ consecutive spike patterns, e.g. $\cO(\omega)=\cO(\bloc{0}{R-1})$.  We consider here that 
observables do not depend explicitly on time (\textit{time-translation invariance of observables}).
As a consequence, for any time $n$, $\cO(\bloc{0}{R-1})=\cO(\bloc{n}{n+R-1})$ whenever $\bloc{0}{R-1}=\bloc{n}{n+R-1}$.

\sssu{Potential}

A function of the form:

\begin{eqnarray}\label{H}
\H : \Omega &\rightarrow & \mathds{R}, \notag \\
\omega &\mapsto & \sum_{k=1}^N \beta_k \cO_k.
\end{eqnarray}
%
is called a \textit{potential}, where the coefficients $\beta_k$ are finite\footnote{Thus, we do not consider here hard core potentials with forbidden configurations.} real numbers. The range of the potential is the maximum of the range of the observables $\cO_k$.

\sssu{Variational principle}

Fix a potential $\H$ as in (\ref{H}). Assume that it has finite range $D$. \footnote{The variational principle still holds if the range is infinite and its variation (\ref{Variation}) decays sufficiently fast with $m$, typically exponentially \cite{ruelle:78,bowen:75,chazottes-keller:11}.}

In this case, a Gibbs distribution $\mu$ obeys the following variational principle:
\beq\label{VarPrinc}
\p{\H}=\sup_{\nu \in \cM_{inv}} \pare{h\bra{\nu} \, + \, \nu\bra{\H}}=
h\bra{\mu} \, + \, \moy{\H},
\eeq
where $\p{\H}$ is called the \textit{topological pressure}, $\nu\bra{\H}=
 \sum_{k=1}^N \beta_k \nu\bra{\cO_k}$ is the average value of
$\H$ with respect to the probability $\nu$ and $\cM_{inv}$ is the set of time-translation invariant probability measures on $\Omega$. We use the notation $\nu(f)$ for the average of a function $f$ instead of $<f>$ used in statistical physics or $\mathds{E}_{\nu}(f)$ used in probability theory. Note that Observables and Gibbs potentials are random functions that acts on the set of raster plots $\Omega$.

Looking at the second equality, the variational principle  (\ref{VarPrinc}) selects, among all possible probabilities $\nu$, a unique one, the Gibbs distribution, realizing the supremum.
A variant of this principle holds when the average value of observables $\cO_k$ is constrained to a value $C_k$, fixed e.g. by experimental observations. In this case $\nu\bra{\H}$ becomes $\sum_{k=1}^N \beta_k C_k$
if the average value of all observables $\cO_k$ is constrained. In this case the variational principle (\ref{VarPrinc}) reduces to maximizing the entropy on the set of measures $\nu \in \cM_{inv}$ such  that $\nu\bra{\cO_k} = C_k$. Then, one is lead to a classical Lagrange multipliers problem where the $\beta_k$s are the Lagrange multipliers. This is the classical approach introduced by Jaynes \cite{jaynes:57}. In this setting (\ref{VarPrinc}) signifies: ``maximizing the entropy given the information that we have of the system" i.e. the observed average value of the observables  $\cO_k$ is $C_k$.

\sssu{Topological pressure}

The topological pressure is the formal analogue of free energy density. It has the following properties:

\begin{itemize}
\item $\p{\H}$ is a log generating function of  cumulants. We have:

\beq\label{AvOk}
\frac{\partial \p{\H}}{\partial \beta_k} =
\mu\bra{\cO_k}.
\eeq
and
\beq\label{d2Pbeta}
\frac{\partial^2 \p{\H}}{\partial \beta_k \partial \beta_l} = \frac{\partial \moy{\cO_k}}{\partial \beta_l} =
\sum_{n=0}^{+\infty} C_{\cO_k \cO_l}(n),
\eeq
where $C_{\cO_k\,\cO_l}(n)$

 $$
 C_{\cO_k\,\cO_l}(n)=\moy{\cO_k \,\cO_l \circ \sigma^n}
 \, - \,  \moy{\cO_k} \moy{\cO_l},
 $$
is  the correlation function between the two observables $\cO_k$ and $\cO_l$ at time $n$ and $\sigma$ is the time shift operator.
Note that correlation functions decay exponentially fast whenever $\H$ has finite range. So that $\sum_{n=0}^{+\infty} C_{\cO_k\,\cO_l}(n) < +\infty$.

Eq. (\ref{d2Pbeta}) characterizes the variation in the average value of
$\cO_k$ when varying $\beta_l$ (linear response).
The corresponding matrix is a susceptibility matrix. It controls the  Gaussian fluctuations of observables around their mean (central limit theorem)
 \cite{ruelle:78,parry-pollicott:90,chazottes-keller:11}.

\item 
$\cP(\H)$ is a convex function of $\bbeta$. 

\item Define:
\beq\label{Zn}
Z_n =  \sum_{\bloc{0}{n}} e^{\H\pare{\bloc{0}{n}}}.
\eeq
The topological pressure obeys:
$$
\cP(\H) = \lim_{n \to +\infty} \frac{1}{n} \log Z_n,
$$
and is analogous to a thermodynamic potential density (free energy, free enthalpy, pressure). 
\end{itemize}

\nid\textbf{Remark 1} For $D>0$ one cannot write the Gibbs distribution in the form:
\beq\label{mu_spatial}
\moy{\bloc{0}{n}}=\frac{1}{Z_n}e^{\H\bra{\bloc{0}{n}}}.
\eeq
It only obeys:
$\exists A, B >0$ such that, for any block 
$\bloc{0}{n}$
$$
  A \leq \frac{\moy{\bloc{0}{n}}}{e^{-(n-D+1)\cP(\H)} e^{\H(\omega_0^n)}} \leq B.
$$
This is actually the definition of Gibbs distributions in ergodic theory \cite{chazottes-keller:11}.

\sssu{Markov chain}

The choice of the potential (\ref{H}), i.e. the choice of a set of observables, fixes the restrictions for the statistical model. A normalization procedure allows to find a normalized potential $\phi$ equivalent\footnote{Two potentials are said ``equivalent" or cohomologous if and only if they correspond to the same Gibbs distribution \cite{jenkinson-etal:05}.} to $\H$ from which the transition probabilities are constructed. This defines  an homogeneous Markov chain whose invariant measure is the Gibbs distribution associated with $\H$. It is constructed as follows.

\sssu*{Transition matrix}
 Consider two spike blocks $w_1,w_2$ of range $D\geq 1$. The transition
$w_1 \to w_2$ is \textit{legal} if $w_1$ has the form $\omega(0)\bloc{1}{D-1}$ and $w_2$ has the form $\bloc{1}{D-1}\omega(D)$. The vectors $\omega(0),\omega(D)$
 are arbitrary but the block  $\bloc{1}{D-1}$ is common.  Here is an example of a legal transition : 
$$
\tiny{w_1 =\left[
\begin{array}{ccc}
0&0&1\\
0&1&1\\
\end{array}
\right] 
};
\,
\tiny{w_2 =\left[
\begin{array}{ccc}
0&1&1\\
1&1&0\\
\end{array}
\right]
}
.$$
 Here is an example of a forbidden transition
$$
\tiny{w_1 =\left[
\begin{array}{ccc}
0&0&1\\
0&1&1\\
\end{array}
\right] 
};
\,
\tiny{w_2 =\left[
\begin{array}{ccc}
0&1&1\\
0&1&0\\
\end{array}
\right]
}
.$$

Any block $\bloc{0}{D}$ of range $R=D+1$ can be viewed as a legal transition from the block $w_1=\bloc{0}{D-1}$ to the block 
$w_2=\bloc{1}{D}$ and in this case we write $\bloc{0}{D} \sim w_1w_2$.  \\

The \textit{transfer matrix} $\cL$ is defined as:
\beq\label{transfermatrix}
\cL_{w_1,w_2}=   
\left\{
\begin{array}{lll}
 e^{\H(\bloc{0}{D})}
\quad &\mbox{if} \quad
w_1, w_2 &\mbox{is legal with } \bloc{0}{D} \sim w_1w_2   \\
0, \quad &\mbox{otherwise}.
\end{array}
\right. .
\eeq

\sssu*{Perron-Frobenius theorem} From the matrix $\cL$ the transition matrix of a Markov chain can be constructed. 
Since  $\H(\bloc{0}{D})>-\infty$,  $e^{\H(\bloc{0}{D})} > 0$ for each legal transition. As a consequence of the 
Perron-Frobenius theorem \cite{gantmacher:66,seneta:06}, $\cL$ has a unique real positive eigenvalue $s_\bbeta$, strictly larger than the modulus of the other 
eigenvalues (with a positive gap), and with associated right, $\rpf$, and left, $\lpf$, eigenvectors: $\cL\rpf=\sb\rpf, \, \lpf\cL=\sb \lpf$.
%

The following holds:

\bit
\item These eigenvectors have strictly positive entries $\rpfc{.}>0$, $\lpfc{.}>0$, functions of blocks of range $D$. They can be chosen so that the scalar product 
$\scal{L,R}=1$. 

\item We have:
\beq\label{Pres_vs_s}
\cP(\H)=\log \sb.
\eeq

\item The following potential: 
\beq\label{Phi_norm}
\phi(\bloc{0}{D})=\H(\bloc{0}{D}) - \Gb(\bloc{0}{D})
\eeq
with:
\beq\label{G}
\Gb(\bloc{0}{D}) = \log \rpfc{\bloc{0}{D-1}} - \log \rpfc{\bloc{1}{D}}+\log s_\bbeta,
\eeq
is equivalent to $\H$ and normalized. It defines a family of transition probabilities:
\beq\label{Prob_cond}
\Probc{}{\omega(D)}{\bloc{0}{D-1}} \deq e^{\phi(\bloc{0}{D})}>0.
\eeq

\item These transition probabilities define a Markov chain which admits a unique invariant probability:
\beq\label{mu_inv}
\mu(\bloc{0}{D-1})=\rpfc{\bloc{0}{D-1}}\lpfc{\bloc{0}{D-1}}.
\eeq
which is the Gibbs distribution satisfying the variational principle (\ref{VarPrinc}).

\item  It follows that the probability of blocks of depth $n \geq D$ is:
\beq\label{P_large_blocks}
\moy{\bloc{0}{n}} = \frac{e^{\H\pare{\bloc{0}{n}}}}{\sb^{n-D+1}} 
\rpfc{\bloc{n-D+1}{n}}\lpfc{\bloc{0}{D-1}}.
\eeq

\item In the case $D=0$ the Gibbs distribution reduces to (\ref{mu_spatial}).  One can indeed easily show that:

$$ \exp \Gb=\sb = \sum_{\omega(0)} \,  e^{\H(\omega(0))}=Z_\bbeta,
$$

 Additionally, since spike patterns occurring at distinct time are independent in the $D=0$ case, $Z_n$ in (\ref{Zn}) can be written as $Z_n = Z_\bbeta^n$ so that $\cP(\H)=\log Z_\bbeta$.

\item \emph{In the general case of spatio-temporal constraints}, the normalization requires the consideration of normalizing function $\Gb$ \textit{depending as well on the
 blocks $\bloc{0}{D}$}.
Thus, in addition to function $\H$ normalization introduces a second \textit{function} of spike blocks.  
This increases consequently the complexity of Gibbs potentials and Gibbs distributions compared to the spatial ($D=0$) case where $\Gb$ reduces to a constant. 
\eit

\sssu{Examples}

We give here a few examples of Maximum Entropy Gibbs distributions, found in the literature.

\begin{itemize}

\item\textit{Bernoulli model.} Here only firing rates of neurons are constrained. The potential has the form:
$$\H(\omega(0)) = \sum_{i=1}^N \beta_i \omega_i(0).$$
This is a memory-less model, where transitions probabilities are given by neuron firing rates $\lambda_i=\frac{e^{\beta_i}}{1+e^{\beta_i}}$. The Gibbs distribution has the form:
\beq\label{Bernoulli}
\moy{\bloc{m}{n}} = \prod_{l=m}^n \, \prod_{k=1}^{N} \, 
\lambda_k^{\omega_k(l)} \, (1-\lambda_k)^{1-\omega_k(l)},  
\eeq
This is thus a product probability where neurons are independent.

\item\textit{Ising model.} This model was introduced by Schneidman et al \cite{schneidman-berry-etal:06} for retina spike train analysis. Here, 
firing rates and instantaneous pairwise synchronisation probabilities are constrained. The potential has the form:
$$\H(\omega(0)) = \sum_{i=1}^N \beta_i \omega_i(0) + 
\sum_{i,j=1}^N \beta_{ij} \omega_i(0) \, \omega_j(0).$$
This is a memory-less model where the Gibbs distribution has the classical form (\ref{mu_spatial}).

\item\textit{Extended spatial Ising model.} A natural extension of Ising model has been proposed by Ganmor et al \cite{ganmor-segev-etal:11a}, where triplets and more general synchronous spike events are considered. The potential has the form:
$$\H(\omega(0)) = \sum_{i=1}^N \beta_i \omega_i(0) + 
\sum_{i,j=1}^N \beta_{ij} \omega_i(0) \, \omega_j(0)+
\sum_{i,j,k=1}^N \beta_{ijk} \omega_i(0) \, \omega_j(0)\, \omega_k(0) + \dots$$
This is a memory-less model where the Gibbs distribution has the classical form (\ref{mu_spatial}).

\item\textit{Spatio temporal Ising model.} In \cite{marre-boustani-etal:09} Marre et al considered a spatio-temporal extension of the Ising model where the potential has the form:
$$\H(\bloc{0}{1}) = \sum_{i=1}^N \beta_i \omega_i(0) + \sum_{i,j=1}^N \beta_{ij} \omega_i(0) \, \omega_j(1).$$
Here spatio-temporal pairs with memory depth $1$ are considered. Although the Gibbs distribution has not the form
(\ref{mu_spatial}), the authors use an approximation of the exact distribution by this form, based on a detailed balance assumption. They applied this model for spike train analysis in the cat parietal cortex.

\item \textit{General Spatio temporal  model.} General models of the form (\ref{H}) have been considered in \cite{vasquez-marre-etal:12,cessac-palacios:12,nasser-marre-etal:12} for the analysis of retina spike trains. A C++ implementation of methods for fitting spatio-temporal models from data is available at \url{http://enas.gforge.inria.fr/v3/}.  

\end{itemize}

\sssu{Applications}

The maximum entropy principle has been used by several authors \cite{schneidman-berry-etal:06,shlens-field-etal:06,tang-jackson-etal:08,yu-huang-etal:08,ohiorhenuan-mechler-etal:10,schneidman-berry-etal:06,ganmor-segev-etal:11a,ganmor-segev-etal:11b} for Multi-electrode Arrays (MEA) spike train analysis. Efficient methods have been designed to estimate the parameters of the potential, in the spatial case \cite{dudik-phillips-etal:04} (Broderick et al., 2007) and in the spatio-temporal case \cite{ nasser-marre-etal:12}.
 
 This approach, grounded on statistical physics, attempts to find a generic model for spike statistics based on a potential of the form (\ref{H}), where the observables and their related $\beta$ parameters summarize ``effective interactions" between spikes. Behind this approach exists, we believe,  a physicists ``dream": inferring, from data analysis, the equivalent of the equation of states existing in thermodynamics; that is, summarizing the behaviour of a big neuronal system by a few canonical variables (analogous e.g. to temperature, pressure, volume in a gas). To our opinion, recent remarkable investigations to exhibit critical phenomena in retina spike train statistics are part of this project (Tka\v{c}ic et al., 2006,2009)  \\

The main advantage of this approach is the possibility of constructing different statistical models based on a priori hypotheses
on the most statistically significant events  (single spikes, pairs, triplets, and so on). As such, it allows to consider arbitrary forms of spatio-temporal correlations. But this strength is also a weakness. Indeed, the possible forms of potentials are virtually infinite and obviously, in the setting of neuronal dynamics, one does not have the equivalent of mechanics or thermodynamics to construct the potential from general principles. 

Finally, this approach only holds for stationary data, a highly questionable assumption as far as data from living systems are concerned.

\subsection{Generalized Linear model}\label{SGLM}

We now consider a second class of Gibbs distributions related to statistical models called Linear-Nonlinear (LN) model and Generalized Linear Model (GLM)
 \cite{brillinger:88, mccullagh-nelder:89,simoncelli-paninski-etal:04, paninski:04,truccolo-eden-etal:05,pillow-paninski-etal:05,pillow-shlens-etal:08,pillow-ahmadian-etal:11,pillow-ahmadian-etal:11b}.   We focus here on the GLM and follow the presentation of Ahmadian et al \cite{pillow-ahmadian-etal:11}. 

\sssu{Conditional intensities}

GLMs are commonly used statistical methods for modeling the relationship between neural population activity and presented stimuli. Let $x \equiv x(t)$ be a time-dependent stimulus. In response to $x$ the network emits a spike train response $r$. This response does not only depend on $x$, but also on the network history of spiking activity. The GLM (and LN) assimilate the spike response $r$ as an inhomogeneous point process: the probability that neuron $k$ emits a spike between $t$ and $t+dt$ is given by $\lambda_k(t \mid \texttt{H}_t) \, dt$, where $\lambda_k(t \mid \texttt{H}_t)$ is called ``conditional intensity" and  $\texttt{H}_t$ is the history of spiking activity up to time $t$.  
In the GLM  this function is given by:
\beq\label{lambdaGLM}
\lambda_k(t \mid \texttt{H}_t) = f\pare{b_k + (K_k \ast x)(t) + \sum_{j} (H_{kj} \ast r_j)(t)},
\eeq
where:
\bit
\item $f$ is a non linear function (an exponential or a sigmoid);

\item $b_k$ is some constant fixing the baseline firing rate of neuron $k$;

\item $K_k$ is a causal, time-translation invariant, linear convolution kernel that mimics a linear receptive field of neuron $k$;  

\item $\ast$ is the convolution product;

\item $H_{kj}$ is the memory kernel that describes possible excitatory or inhibitory post spike effects of the $j^{th}$ observed neuron on the $k^{th}$. As such, it depends on the past spikes, hence on $\omega$.
The diagonal
components $H_{kk}$ describe the post spike feedback of the neuron to itself,
and can account for refractoriness, adaptation and burstiness depending on their shape;
\item $r_j$ is the spike train of neuron $j$: $r_j(t)=\sum_{r \geq 1} \delta(t-t_j^{(r)})$, where $t_j^{(r)}$ is the time of the $r^{th}$ spike of $j^{th}$ neuron. 
\eit

The spike response has a history dependent structure that makes Poisson models inappropriate. Point processes affords for history dependence and generalizes Poisson process. 
A point process can be completely characterized by its conditional intensity function.

$$\lambda_k(t \mid \texttt{H}_t)=\lim_{\Delta t \rightarrow 0}\frac{\mathds{P}(\Delta N_{[t +\Delta t)=1} \mid \texttt{H}_t)}{\Delta t},$$
where $N_{[t +\Delta t)}$ is the counting process that gives the number of spikes occurring in the interval $[t +\Delta t)$. Choosing $\Delta t$ to be a sufficiently small time interval $\sim 1ms$, the probability of firing more than one spike is negligibly small compared to the probability of firing one spike.
This assumption is biophysically plausible because neurons have refractory period. Therefore:

$$\mathds{P}(\text{spike in } [t +\Delta t)  \mid \texttt{H}_t)  \approx \lambda_k(t \mid \texttt{H}_t)\Delta t.$$

Here $\lambda_k(t \mid \texttt{H}_t)$ is defined in continuous time, and spikes are discrete events. If we discretize the time to make the spikes emitted by the point process belong to a single bin, we have:

$$\mathds{P}(\omega_k(n)=1  \mid \texttt{H}_{n-1})  \approx \lambda_k(n \mid \texttt{H}_{n-1})\Delta t := p_k(n)$$

\sssu{Conditional independence}

 The GLM postulates that, given the history $\texttt{H}$ and stimulus $x$, neurons are independent (\textit{conditional independence upon the past and stimulus}). In the context of transition probabilities defined on section \ref{SPTrans}, the response at time $n$ is a spiking pattern $\omega(n)$ while the history is the spike activity $\texttt{H}$. As a consequence of the conditional independence assumption the probability of a spike pattern follows a Bernoulli process:

\beq\label{PGLM}
\Probc{n}{\omega(n)}{\sif{n-1}} = \prod_{k=1}^N p_k(n)^{\omega_k(n)} (1- p_k(n))^{1-\omega_k(n)}.
\eeq


\sssu{Gibbs distribution}

Transition probabilities are strictly positive whenever $0 < p_k(n) <1$, for all $k$,$n$. If $f$ is e.g. a sigmoid this holds provided its argument $b_i + (K_i \ast x)(t) + \sum_{j} (H_{ij} \ast r_j)(t)$ remains bounded in absolute value. The continuity of $\lambda$ with respect to $\omega$ holds whenever $f$ is continuous and the memory kernel $H$ is continuous  with respect to $\omega$. This second condition is fulfilled in two cases:

\begin{itemize}

\item $H$ depends on a finite past;

\item $H$ depends on an infinite past, but the memory dependence decays sufficiently fast to ensure continuity. Since $H$ mimics synaptic influence it is typically a sum of $\alpha$-profiles that mimic PSPs (Post Synaptic Potentials). $\alpha$ profiles decay exponentially fast with time, so they match this condition. We come back to this point in section \ref{SGIF}.

\end{itemize}

The Gibbs potential associated with (\ref{PGLM}) is:
\beq\label{phiGLM}
\phi_n(\omega)=\sum_{k=1}^N \pare{\omega_k(n)\log p_k(n) + (1-\omega_k(n))(1-p_k(n))},
\eeq
%

It is normalized by definition.

\sssu{Applications}

This model has been applied in a wide variety of experimental settings \cite{brillinger:92,chichilnisky:01,theunissen-david-etal:01,brown-barbieri-etal:03,paninski-fellows-etal:04,truccolo-eden-etal:05,
pillow-shlens-etal:08}. Efficient methods has been designed to estimate the parameters \cite{pillow-ahmadian-etal:11}. 

To us, the main advantages of the GLM are:

\begin{itemize}
\item The transition probability is known (postulated) from 
the beginning and does not require the heavy normalization  (\ref{Phi_norm}) imposed by potentials of the form (\ref{H}); 

\item The model parameters have a  
neurophysiological interpretation, and their number grows at most as a power law in the number of neurons.

\item It has good decoding performances 

\item It holds for non stationary data.
\end{itemize}

Its main drawbacks are:

\begin{itemize}
\item It postulates an ad hoc form for the transition probability of the stochastic process;

\item It uses a quite questionable assumption of conditional independence: neurons are assumed independent at time $n$ when the past is given. On the opposite,  the maximal entropy principle does not require this assumption.

\item To us, the biophysical interpretation of the parameters $H_{kj}$ is unclear. Do they correspond to ``real" connectivity ? ``functional" connectivity ? 

\end{itemize}

\subsection{Integrate and Fire neural networks}\label{SGIF}

The previous examples were mainly developed for data analysis: one speculates a form for transitions probabilities, performs parameters fitting, and then uses the model to decode or to extrapolate the statistics of complex events. Here we start from a different point of view asking the following questions: Can we have a reasonable idea of what could be the spike train statistics studying a neural network model? Do Gibbs distribution arise in these models ? What is the shape of the potential ? We focus here on a model proposed in \cite{cessac:11a,cessac:11b,cofre-cessac:12} where these questions have been answered. 

\sssu{Model}
The integrate-and-fire model remains one of the most ubiquitous model for simulating and analyzing the dynamics of neuronal circuits. Despite its simplified nature, it captures some of the essential features of neuronal dynamics.  Denote $V(t)$ the membrane potential vector with entries $V_k(t)$. The continuous-time dynamics of $V(t)$ is defined as follows.
Fix a real variable $\theta>0$ called ``firing threshold". 
For a fixed time $t$, we have two possibilities:

\begin{enumerate}

\item Either $V_k(t) < \theta$, $\forall k=1, \dots,N$. This corresponds to \textit{sub-threshold dynamics}.

\item Or, $\exists k$, $V_k(t) \geq \theta$. Then, we speak of \textit{firing dynamics}.
\end{enumerate}
 
The model proposed here is an extension of the conductance based Integrate-and-Fire neuron model introduced in \cite{rudolph-destexhe:06}. The model-definition follows the presentation given in \cite{cessac-vieville:08,cessac:11b}. Neurons are considered as points, with neither spatial extension nor biophysical structure (axon, soma, dendrites). 
Dynamics is ruled by a set of stochastic differential equations where parameters, corresponding to chemical conductances, depend on the action potentials emitted in the past by the neurons. In this way, the dynamical system defined here is ruled both by continuous and discrete time dynamical variables.

\sssu*{Subthreshold dynamics}
It is defined by:
\beq\label{sub_threshold}
C_k \frac{dV_k}{dt} = -g_{L,k}(V_k-E_{L})-\sum_j g_{kj}(t,\omega)(V_k-E_j) + \sum_j \gap{kj} \left( V_j - V_k\right)+I_k(t),
\eeq
where:
\begin{itemize}
\item $C_k$ is the membrane capacity of neuron $k$;

\item $I_k (t) = i_k^{(ext)}(t) + \sigma_B \, \xi_k (t)$ is a current when a time-dependent part $i_k^{(ext)}(t)$ (stimulus) and stochastic part $\sigma_B \, \xi_k (t)$ where $\xi_k (t)$ is a white noise and $\sigma_B$ controls the noise intensity;

\item $g_{L,k}$ is the leak conductance and $E_{L} < 0$ the leak Nernst potential;

\item $\gap{kj}$ mimics electric conductance (gap junctions) between neurons $j$ and $k$; these are passive and symmetric conductances; 

\item  the term
\beq\label{gkj}
g_{kj}(t,\omega)=G_{kj} \sum_{r: \tf{j}{r}(\omega) <t} \alpha_{kj} \left( t-\tf{j}{r}(\omega)\right),
\eeq
mimics the conductance of the chemical synapse $j \to k$,
where:
\beq\label{alpha}
\alpha_{kj}(t) = \frac{t}{\tau_{kj}}e^{-\frac{t}{\tau_{kj}}}H(t),
\eeq
mimics a PSP, $H(t)$ is the Heaviside function (that mimics causality); $\tf{j}{r}(\omega)$ is the $r^{th}$ spike emitted by neuron $j$ in the raster $\omega$,
therefore $g_{kj}(t,\omega)$ depends on the whole spike history; $E_j$ is the reversal potential of the chemical synapse $j \to k$.
 
\end{itemize}

\sssu*{Firing dynamics and reset}

If, at time $t$, some neuron $k$ reaches its firing threshold $\theta$, $V_k(t)=\theta$, then this neuron emits a spike.
To conciliate the continuous time dynamics of membrane potentials and the discrete time dynamics of spikes we define the spike and reset as follows.

\bit

\item The neuron membrane potential $V_k$ is reset to $0$
at the \textit{next integer time}  after $t$. 

\item A spike is registered at time $[t]+1$ where $\ent{t}$ is the integer part of $t$. This allows us to represent spike trains as events on a discrete time grid. It has the drawback of artificially synchronizing spikes coming from different neurons, in the deterministic case 
\cite{cessac-vieville:08,kirst-timme:09}. However, the presence of noise in membrane potential dynamics eliminates this synchronization effect. 

\item Spikes are separated by a time scale $\tau_{sep}>0$ which is a multiple of $\delta$ (thus an integer). 

\item Between $[t]+1$ and $[t]+\tau_{sep}$ the membrane potential $V_k$ is maintained to $0$ (refractory period).
From time $[t]+\tau_{sep}$ on, $V_k$ evolves according to (\ref{sub_threshold}) until the next spike.

\item When the spike occurs (at time $[t]+1$), the raster $\omega$ as well conductances $g_{kj}(t,\omega)$ are updated. 
 
\eit

\sssu{Main results}

This model has several variants: discrete time \cite{cessac:11a}; continuous time with chemical synapses  \cite{cessac:11b} and continuous time with chemical and electric synapses \cite{cofre-cessac:12}. We list here the main results concerning spike statistics and Gibbs distributions.

\ben

\item Whatever the values of the parameters the model admits a unique Gibbs distribution in the general sense given in section \ref{SGibbsdef}.

\item When the noise is weak and \textit{without gap junctions}, the normalized Gibbs potential can be explicitly computed. It takes the form:
\beq\label{phiGIF}
\phi_n(\omega)=\sum_{k=1}^N \pare{\omega_k(n)\log \lambda_k(n) + (1-\omega_k(n)\log(1-\lambda_k(n))},
\eeq
where
\beq\label{PGIF}
\lambda_k(n) = f\pare{
b_k(n-1,\omega)
+
\Phi^{(ext)}_k(n-1,\omega)
+
\Phi^{(syn)}_k(n-1,\omega)
},
\eeq
where:
\bit
\item $f$ is a sigmoid function;

\item  $b_k(n-1,\omega)$ is a function depending on the threshold value, the leak Nernst potential, and on the integrated noise, integrated from the last time where has been reset (depending on $\omega$) up to time $n-1$;

\item $\Phi^{(ext)}_k(n-1,\omega)$ corresponds to the integrated effects of the external current $i^{(ext)}_k$ on the membrane potential;

\item the term $\Phi^{(syn)}_k(n-1,\omega)$ corresponds to the integrated effects of chemical synapses on the membrane potential.

\eit
\item Eq. (\ref{phiGIF}), expresses that in this case, neurons are \textit{conditionally independent upon the past}.  

\item In this conductance-based model, conductances depend on the past via (\ref{gkj}). One can consider as well a current-based model where conductances are fixed and current depend on the past spikes. In this case, the terms $\Phi^{(ext)}_k(n-1,\omega)$ and $\Phi^{(syn)}_k(n-1,\omega)$ can be written as convolutions and one recovers a potential with a form analogous to (\ref{phiGLM}).

\item \emph{In the general case (gap junctions)}, neurons are \textit{not conditionally independent}. Gap junctions induce a coupling effect which does not allow any more the factorization (\ref{phiGIF}) of the potential.  

\item Correlations (pairwise and higher order) are mainly due to chemical synapses and gap junctions. Additional correlations can also
be induced by the stimulus using e.g. a current $i^{(ext)}$
where time fluctuations of $i_k^{(ext)}$ are correlated with
$i_j^{(ext)}$. But these are extra-correlations that disappear when the stimulus is removed, whereas the dynamical correlations remains. 

\item The potential has an infinite range (infinite memory). However, thanks to the exponential decay of the alpha profile, one can show that the potential is continuous.
This allows to propose Markovian approximation of the Gibbs distribution where the exact potential is replaced by a potential with a finite range \cite{cessac:11a,cessac:11b}.
\een

\sssu{Applications}

What do we finally learn from the study of this model ?

\bit
\item We have a positive answer to the existence of Gibbs distributions in neural networks models.

\item An explicit form for the potential is known in specific cases.
%

\item The form (\ref{PGIF}) actually also fits with maximum entropy model, in the stationary case, as shown in section \ref{SRel}.

\item In this model the origin of correlations is essentially due to dynamics, not to the stimulus.

\item Gap junctions play here a central role in the structure of dynamical correlations and dependence of dynamics upon history (see \cite{cofre-cessac:12} for more details).

\item The analysis holds for non stationary data.
\eit

Considering potential uses of this study to fit real data the main criticism is:

\bit

\item This is a model. Is it sufficient to describe real neural networks ? For example, its application to retina data is controversial since it considers only spiking cells (that mimics ganglion cells), but retina has also non firing cells like most amacrine and bipolar cells.

\item In the general case there is no explicit form for the Gibbs potential.

\item Even when there exists an explicit form for the potential, it has quite a lot of parameters which can be difficult to fit from data. 
\eit

\ssu{Relations between these approaches}\label{SRel}

In this section, we establish a connection between the three examples of Gibbs distributions considered in sections \ref{SMaxEnt}, \ref{SGLM}, \ref{SGIF}.\\

Consider a family of transition probabilities satisfying the positivity condition (ii) in section \ref{SGibbsdef}, where we furthermore assume that the memory depth is finite and that transition probabilities are time-translation invariant. As stated in section \ref{SMarkov} this define an homogenous Markov chain. The transition probabilities are thus functions 
of blocks $\bloc{0}{D}$ (see section \ref{SPTrans} and the definition of time translation invariance). These functions can then take at most $2^{N(D+1)}$ values. The same holds for
the normalized potential (\ref{phin}). Now, one can prove
that any such function  can be written as:
\beq\label{phiHS}
\phi(\omega) \equiv \phi(\bloc{0}{D}) =
\sum_{l=0}^{L} \phi_l \, \cO_l(\omega),
\eeq 
with $L=2^{N(D+1)-1}$ and $\cO_l$ is an observable  (\ref{DefObs}) where the time index ranges from $0$ to $D$. The index $l$ parametrizes an enumeration of all possible observables with $N$ neurons and $D+1$ time steps,
where $m_0$ is the constant observable $\cO_0=1$, and so on.

Now, using the positivity condition and the results in section \ref{SMaxEnt} one can
show that any family of stationary transition probabilities with memory depth $D$ can be associated with a potential of the form (\ref{H}). The correspondence is actually unique.
This is a straightforward application of the celebrated Hammersley-Clifford theorem \cite{hammersley-clifford:71,besag:74,moussouris:74,clifford:90}.\\

An immediate consequence of this result is that, \emph{in the stationary case with finite memory}, the GLM potential (\ref{lambdaGLM}) and the Integrate and Fire (\ref{phiGIF}) 
correspond to a Maximum Entropy model with a potential of the form (\ref{H}). The parameters $\beta_
k$ in (\ref{H}) are then 
nonlinear functions of the parameters in (\ref{phiGLM}) or in (\ref{phiGIF}) (see \cite{cessac:11a}). As a consequence, some of these parameters are redundant: there are a priori $2^{N(D+1)}$ non vanishing parameters $\beta_k$ while there are quite less parameters in the GLM or in the  Integrate and Fire (of order $N^2$).
However, the GLM assumes conditional independence between neurons, while the maximum entropy approach is precisely used to take care of (pairwise and higher order) correlations between neurons. In this sense it is more general.

In the non stationary case one can no more apply the maximum entropy principle (entropy is not defined). However, in the case where statistics depends on time on a slow time scale (compared to spike characteristic time scale) one can use a quasi-static approach where the parameters $\beta_k$
in (\ref{H}) vary slowly in time \cite{roudi-hertz:11} (Tyrcha et al., 2012). On the opposite,
the GLM allows to consider non stationary data with efficient results \cite{pillow-ahmadian-etal:11,pillow-ahmadian-etal:11b}. 

The IF model contains both maximum entropy models and GLM. It has a maximum entropy Gibbs distribution in the stationary case, and it reduces to GLM upon several simplifications. In its more general form it allows the consideration of non stationarity and does not rely on the conditional independence assumption. Unfortunately, its generality is a weakness since an explicit form for the potential is not known yet in the general case.

\section{Conclusion}

In this paper we have argued that Gibbs distribution considered in a fairly general sense could constitute generic statistical models to fit spike trains data. The example of the Integrate and Fire model suggests that such distribution could be also defined for more  elaborated neural networks models (FitzHugh-Nagumo or Hodgkin-Huxley). In particular, the
existence and uniqueness of a Gibbs measure holds whenever there is continuity with respect to a raster, with a sufficiently fast decay of the variation (\ref{Variation}) \cite{fernandez-maillard:05}. As shown in \cite{cessac:11a,cessac:11b,cofre-cessac:12} this property is ensured when interactions between neurons decay exponentially fast. This is typically the case for chemical synapses where the PSP (\ref{alpha}) decays exponentially fast with time.

The interest of proposing Gibbs distribution constructed from neural network models is multiple. The model mimics a neurophysiological structure where interactions between neurons, stimuli, and biophysical parameters are well identified. As a consequence the model-parameters can be easily interpreted. Thus, the role of each specific biophysical parameter on spike statistics can be easily analysed. Also, the potential obtained this way is already normalized, while e.g. maximum entropy principle requires a complex procedure to achieve normalisation. Finally, in this context, it is possible to study the effect of a time-dependent stimulus on spike statistics.

However, this approach, to be efficient requires (i) to have an analytical form for the potential; (ii) to be able to fit the many parameters of a non linear problem. This is yet far from being achieved.

\bigskip

\textbf{Acknowledgments}
This work was supported by the French ministry of Research and University of Nice (EDSTIC), INRIA, ERC-NERVI number 227747, KEOPS ANR-CONICYT and European Union Project $\#$ FP7-269921 (BrainScales), Renvision $\#$ 600847 and Mathemacs  $\#$ FP7-ICT-2011.9.7.
\bibliographystyle{plain}	
\bibliography{biblio,odyssee}	 

\end{document}